\documentclass[prl,showpacs,twocolumn,floatfix]{revtex4}
\usepackage{graphicx}
\usepackage{bm}

\begin{document}


\title{ 
 High momentum response of liquid $^{\bm 3}$He  
}
\author{F. Mazzanti,$^1$ A. Polls,$^2$ J. Boronat,$^3$ and J. Casulleras$^3$}
\affiliation{ $^1$ Departament d'Electr\`onica, Enginyeria i Arquitectura
  La Salle,  
  Pg. Bonanova 8, Universitat Ramon Llull, 
  E-08022 Barcelona, Spain}
\affiliation{ $^2$ Departament d'Estructura i Constituents de la
  Mat\`eria, 
  Diagonal 645, Universitat de Barcelona, 
  E-08028 Barcelona, Spain\\}
\affiliation{$^3$ Departament de F\'\i sica i Enginyeria Nuclear,
  Universitat Polit\`ecnica de Catalunya, 
  Campus Nord B4-B5, E-08034 Barcelona, Spain}

\date{\today}


\begin{abstract}
A final-state-effects formalism suitable to analyze the high-momentum 
 response of Fermi liquids is presented and used to study the dynamic structure
function  of liquid $^3$He. The theory, 
developed as a natural extension of the Gersch-Rodriguez formalism,
incorporates the Fermi statistics explicitely through a new additive term
which depends on the semi-diagonal two-body density matrix. The use of a
realistic momentum distribution, calculated using the diffusion Monte Carlo
method, and the inclusion of this additive correction allows for a good
agreement with available deep-inelastic neutron scattering data.
\end{abstract}


\maketitle


Inelastic neutron scattering is the most efficient tool to explore the
structure and dynamics of quantum liquids $^4$He and $^3$He since the dynamic structure
function $S(q,\omega)$ is readily obtained from the double differential
scattering cross-section~\cite{glyde1}. The range of momenta $q$
transferred to the system determines the kind of microscopic information
that can be extracted. The most interesting regimes correspond to low and
high $q$'s. At low $q$, the scattering data allows for the determination of
the low-energy excitation spectrum. In the opposite limit, 
known as deep-inelastic neutron scattering (DINS), $q$ is so high that
single-particle properties of the system become accessible.  

It is well known that in the $q \rightarrow \infty$
limit, $S(q,\omega)$ approaches the impulse
approximation (IA). The only ingredient to calculate the response in IA is 
the momentum distribution $n(k)$, a fundamental function in the study of
$^4$He, $^3$He, and the  $^4$He-$^3$He mixture. The boson and fermion quantum
statistics of $^4$He and $^3$He, respectively, introduce  significant
differences in their corresponding momentum distributions. Liquid $^4$He presents
a macroscopic occupation of the zero-momentum state, characterized by its
condensate fraction $n_0$; $n(k)$ of liquid $^3$He, considered as a normal
Fermi liquid, shows a discontinuity at the Fermi momentum $k_{\text F}$~\cite{mom}.
Nowadays, different theoretical calculations of $n(k)$ ranging from variational
theory, based on the (Fermi-)hypernetted-chain equations ((F)HNC),  
to the more exact diffusion Monte
Carlo (DMC) method are in an overall quantitative agreement
\cite{fantoni,pandha,rosati}. However, a
direct comparison with experimental data is not possible due to
instrumental resolution effects (IRE) and, more fundamentally, to final state
effects (FSE). From the theoretical side, the problem is that the IA does not account 
completely for the scattering in
most of the DINS experiments since the transferred momenta are
not high enough. Therefore, FSE which take into account the interactions of the struck 
atom with the medium can not be disregarded.

The search for an unambiguous experimental signature of $n_0$ in
liquid $^4$He using DINS has originated  a great deal of theoretical and experimental
work for the last two decades. At present, theoretical predictions 
\cite{carraro,mazz3,fabro1,rinat,glyde2}
for both the FSE and the IA provide a satisfactory description of the experimental 
measurements,
 with an overall agreement on the value of the  
condensate fraction, $n_0 \sim 9$\% at the equilibrium density. It is worth
noticing that FSE in superfluid $^4$He are enhanced due to $n_0$
and therefore,  even at the highest momenta achieved 
in the laboratory, FSE play a fundamental role. Comparatively, few works
have been devoted
to the analysis of the high-$q$ response  of liquid $^3$He. The
main reasons underlying this situation are, from the experimental side, 
the large neutron absorption of $^3$He, and from the theoretical one, the
difficulties the Fermi statistics of $^3$He introduces in the quantum
many-body calculations. 
The most accurate data have been reported
by Azuah \textit {et al.} \cite{azuah}, and more recently by Senesi 
\textit {et al.}~\cite{senesi}, but only the first one was carried out at the
equilibrium density. FSE in $^3$He have been taken into account
by Moroni \textit {et al.} \cite{fabro1} using the bosonic formalism of
Carraro and Koonin \cite{carraro}. Their results \cite{fabro1} 
show less strength at the peak than the experimental $S(q,\omega)$, 
pointing to possible limitations of the formalism when applied to a Fermi
liquid. On the other hand, 
an analysis of the experimental data based on cumulant expansions 
\cite{glyde1,glyde3} has revealed significant differences between the 
\textit{experimental} and theoretical 
$^3$He momentum distributions at equilibrium density \cite{azuah}.

We present in this letter results for the high-$q$  response of $^3$He using a
theoretical formalism that incorporates explicitly and consistently the
Fermi statistics to the FSE. The inputs required are the
momentum distribution $n(k)$ and the semi-diagonal two-body density matrix  
$\rho_2(\bm{r}_1,\bm{r}_2;\bm{r}_1^\prime,\bm{r}_2)$. Both are obtained
from microscopic theory: $n(k)$ from a DMC
calculation, and $\rho_2$ from variational FHNC. The results
obtained for $S(q,\omega)$ reproduce the experimental data better than
previous estimations, pointing to non-negligible Fermi contributions to 
the FSE.

As long as $^4$He is concerned, most theories introduce FSE as a
convolution in energies, which turn
into an algebraic product in time representation. Hence, FSE are included 
by means of a new function $R(q,t)$ which multiplies $S_{\text{IA}}(q,t)$ 
to obtain the total response, $S(q,t)= S_{\text{IA}}(q,t) R(q,t)$.
This convolutive approach is clearly ambiguous when applied to fermions
because, contrary to the Bose case,
$S_{\text{IA}}(q,t)$ has an infinite number of nodes, a fact that leads to
a singular definition of $R(q,t)$. In order to overcome this serious
drawback that emerges from a direct translation of the FSE theories for
bosons to fermions, we have used an alternative formulation  that
can be considered
a natural extension of Gersch-Rodriguez theory \cite{gersch} to fermionic systems. 
In previous works \cite{mazz1,mazz2}, we have applied this theory to
evaluate FSE
in $^4$He-$^3$He mixtures, but there the fermionic corrections are
much smaller due to the low $^3$He concentrations.

At high $q$, the density-density correlation factor $S(q,t)$ can be well approximated by
\begin{eqnarray}
S(q,t) & = & \frac{1}{N!} e^{i \frac{\omega_q}{v_q} s} \int d\bm{r}_1
\ldots d\bm{r}_N \,
\rho_N(\bm{r}_1,\ldots,\bm{r}_N;\bm{r}_1+\bm{s}) \nonumber \\
& & \times \exp \left[ \frac{i}{v_q}
\sum_{j=2}^N \int_0^s \Delta V(\bm{r}_{1j},\bm{s}) \right ] \label{eq-def2} ,
\end{eqnarray}
with $\omega_q=q^2/2m$, $\bm{v}_q= \bm{q}/m$, $\bm{s}=\bm{v}_q t$, and
$\Delta V(\bm{r},
\bm{u})= V(\bm{r}- \bm{u})-V(r)$. Equation (\ref{eq-def2})  is derived
assuming that the atom struck
 by the neutron recoils in a medium of non-moving $^3$He atoms.
$S(q,t)$  is still hard to evaluate using Eq. (\ref{eq-def2})
since it implies an integration over the complete semi-diagonal $N$-body 
density matrix of the system, which is
essentially the square of the ground-state wave function. 
In $^4$He, a truncated cumulant expansion of Eq.~(\ref{eq-def2}) leads 
to the Gersch-Rodriguez expression for the FSE \cite{gersch}. Recently, this
formalism has proven its efficiency by reproducing $^4$He DINS data
with high accuracy \cite{mazz3}. On the contrary, the nodal structure
of $\rho_N$ in a Fermi system prevents from a straightforward application of 
these methods. 

In order to extend the FSE theory to $^3$He, one introduces an auxiliary 
$N$-body density matrix $\rho_N^{\text B}$ 
corresponding to a system of spinless bosons with the mass, density,
and interatomic potential of $^3$He.   
$\rho_N^{\text B}$ is positive and then it can be used as the starting point of a 
cumulant expansion of the response function. 
In terms of $\rho_N^{\text B}$, a convenient decomposition of $\rho_N$ turns out to be 
\begin{eqnarray}
\lefteqn{\rho_N ( \bm{r}_1, \ldots, \bm{r}_N; \bm{r}_1^{\prime})  =
  \rho_1(r_{11'}) \left[
\frac{1}{\rho_1^{\text B}(r_{11'})}  \right . }  \nonumber \\
& & \times \rho_N^{\text B}(\bm{r}_1, \ldots, \bm{r}_N; \bm{r}_1')
\bigg ] 
+\Delta \rho_N(\bm{r}_1, \ldots, \bm{r}_N; \bm{r}_1')
\label{deltarho} \ , 
\end{eqnarray}
$\rho_1^{\text B}(r_{11'})$ being  the one-body density matrix extracted
from $\rho_N^{\text B}$. In  the thermodynamic limit, $\rho_1^{\text
B}(r_{11'})$ factorizes from $\rho_N^{\text B}$, and thus the first term in Eq.
(\ref{deltarho}) corresponds to an artificial  $N$-body density matrix 
containing fermionic correlations between points 1 and
1$^\prime$ only.

Inserting $\rho_N$ (\ref{deltarho}) in $S(q,t)$  (\ref{eq-def2}), the
$^3$He response becomes
 \begin{equation}
S(q,t)= S_{\text{IA}}(q,t) R(q,t) + \Delta S(q,t) \ , 
\label{eq-decom}
\end{equation}
with $S_{\text{IA}}(q,t)$ the exact $^3$He IA, and $R(q,t)$ the
Gersch-Rodriguez FSE function calculated with the 
bosonic semi-diagonal two-body density matrix $\rho_2^{\text
B}(\bm{r}_1,\bm{r}_2;\bm{r}_1')$,
\begin{eqnarray}
R(q,t) & = & \exp \left [ - \frac{1}{\rho_1^{\text B}(r_{11'})} \int d
\bm{r} \rho_2^{\text B} (\bm{r},0;
\bm{r}+\bm{s})  \right . \nonumber \\
& & \times
\left . \left [1 -\exp \left (\frac{i}{v_q} \int_0^s ds' \Delta
V(\bm{r},\bm{s}^\prime)
\right ) \right ] \right ]  \label{eq-fse}  \ .
\end{eqnarray}
The new additive term  $\Delta S(q,t)$ in Eq. (\ref{eq-decom}) is a consequence of
$\Delta \rho_N$ introduced in Eq. (\ref{deltarho}). The leading contribution to 
the FSE at high $q$ depends on the semi-diagonal two-body density
matrix, and then $\Delta \rho_2$ is required for the calculation of $\Delta
S(q,t)$. A  cluster expansion in the framework of the FHNC 
formalism allows for an estimation of $\Delta \rho_2$ according to the
following structure,
\begin{eqnarray}
\frac{1}{\rho}  \Delta \rho_2(\bm{r}_1,\bm{r}_2; \bm{r}_1^\prime) & = & 
\rho_1(r_{11'}) G(\bm{r}_1,\bm{r}_2;\bm{r}_1^\prime) \nonumber \\
& & + \, 
\rho_{\text{1D}}(r_{11'}) F(\bm{r}_1,\bm{r}_2;\bm{r}_1') \label{eq-rho2-decom} \ .
\end{eqnarray}
The form factors $G(\bm{r}_1,\bm{r}_2;\bm{r}_1')$ 
and $F(\bm{r}_1,\bm{r}_2;\bm{r}_1')$ can be expressed in terms of auxiliary 
functions defined in the FHNC theory,
and $\rho_{\text{1D}}(r_{11'})$ is positive everywhere
 and similar to a bosonic one-body density matrix \cite{clark,polls} . 
 Therefore, $\rho_{1D}(r_{11'})$ can be used  as the basis
of a cumulant expansion by simply adding and subtracting it to $\Delta \rho_N$. 
The resulting additive correction $\Delta S(q,t)$ is, to the lowest
order,
\begin{widetext}
\begin{equation}
 \Delta S(q,t)  =   e^{i \omega_q s/v_q} \, \frac{1}{\rho} \,
\rho_{\text{1D}}(r_{11'})     
\left \{ 
\exp \left [ - \frac{1}{\rho_{\text{1D}}(s)} \int d \bm{r} 
\Delta \rho_2(\bm{r},0; \bm{r}+\bm{s})   
 \left [ 1 -\exp \left ( \frac{i}{v_q} \int_0^s ds' \Delta
 V(\bm{r}, \bm{s}^\prime) 
\right ) \right ] \right ] - 1 \right \}     \ .
\label{deltas}
\end{equation}
\end{widetext}
Finally, a Fourier transform of $S(q,t)$ provides the dynamic structure function
$S(q,\omega)$.  Furthermore, the scaling 
properties of the IA, in terms of the West variable \cite{west}
$Y=m\omega/q -q/2$, suggests as usual to write the response $(q/m) S(q,\omega)$
as a Compton profile, 
\begin{equation}
J(q,Y) = \int d Y' J(Y') R(q,Y-Y') + \Delta J(q,Y) \ ,
\label{finalres}
\end{equation}
$J(Y) = 1/(2 \pi^2 \rho) \int_{\mid Y\mid}^{\infty} dk \, k n(k)$ being the IA.

The microscopic functions entering the high-momentum response $J(q,Y)$
(\ref{finalres}) are the one- and the semi-diagonal two-body density matrices of the 
actual system and its  bosonic counterparts.
The most relevant quantity 
 is the one-body density matrix, or equivalently the momentum distribution, which is used 
to evaluate $J(Y)$. We have estimated the $^3$He  momentum distribution  
 using the DMC methodology that has recently proved to be very accurate in
 the calculation of the $^3$He equation of state at zero temperature
 \cite{casu,boronat}. At the equilibrium density $\rho_0=0.273\ \sigma^{-3}$
 ($\sigma=2.556$ \AA), $n(k)$ is well parameterized by 
\begin{equation}
n(x)= \left \{ \begin{array}  {ll}
a_0 - a_3 x^3   &   x \leq 1  \\
(b_0 + b_1 x + b_2 x^2) e^{-b_t x }    &          x > 1
\end{array}   \right . \ ,
\label{modelnk}
\end{equation}
with $x=k/k_{\text F}$, $k_{\text F}$ being the Fermi momentum. The 
set of parameters that best fit $n(k)$ (\ref{modelnk}) is reported in Table I. The
kinetic energy per particle, related to the second moment of $n(k)$, is
12.3 K and the  discontinuity of $n(k)$ 
at the Fermi surface is $Z=0.236$. The value of $Z$,  
which defines the strength of the quasi-particle pole, is rather small,
indicating that the system is strongly correlated.
On the other hand, the tail of the momentum distribution extends up to high momenta
generating significant high-energy wings in $J(q,Y)$. The present $n(k)$ is in 
overall agreement with the DMC one from Ref. \cite{fantoni}.

\begin{table}[b]
\centering
\begin{ruledtabular}
\begin{tabular}{cl}  
a$_0$   &  0.481319  \\
a$_3$   &  0.0842956   \\
b$_0$   &  1.39056    \\
b$_1$   &  0.157930   \\
b$_2$   &  0.0829832   \\
b$_t$   &  2.31398 
\end{tabular}
\end{ruledtabular}
\caption{Parameters of $n(k)$ (\protect\ref{modelnk}) at $\rho_0$.}
\end{table}
 


The semi-diagonal two-body density matrix $\rho_2$, and the auxiliary bosonic 
functions $\rho_1^{\text B}$ and $\rho_2^{\text B}$,  have been
obtained  in the framework of the FHNC and HNC theories using  a
Jastrow-Slater variational wave function. It is well known that this trial
wave function is not accurate enough if the main objective is to get a good
upper-bound to the energy. This is not however the aim of the present
letter. In fact, 
we have shown in previous work that a Jastrow wave function can
efficiently account for the FSE in $^4$He \cite{mazz3}. Certainly,  
short-range correlations, which
dominate the FSE, are already contained in the Jastrow-Slater
approximation. Accordingly, the  diagrammatic
analysis for $^3$He  has been performed at the two-body level, thus making
the analysis easier.

\begin{figure}[b]
\centerline{
\includegraphics*[width=0.8\linewidth]{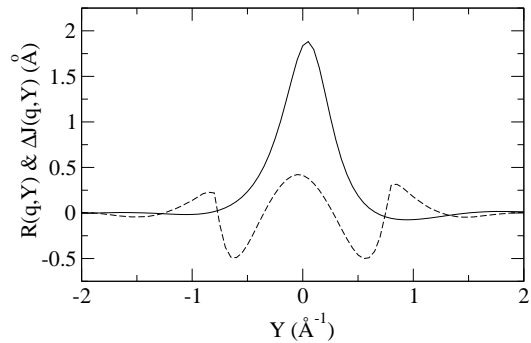}
}
\caption{FSE broadening function $R(q,Y)$ (solid line) and additive correction $\Delta
J(q,Y)$ (dashed line) at $q=19.4$ \AA$^{-1}$ and $\rho_0$. $\Delta J(q,Y)$
is multiplied by a factor 20 to fit into the scale.}
\label{fig-rqy}
\end{figure}

The FSE broadening function $R(q,Y)$ at $\rho_0$ and
a momentum transfer $q=19.4$ \AA$^{-1}$ 
is shown in Fig. \ref{fig-rqy}. This value of $q$ has been used throughout
this work since it corresponds to the momentum reported in the 
experimental data by Azuah \textit{et al.} \cite{azuah}.  
$R(q,Y)$ has been calculated in the bosonic approximation and then its
structure is similar to the FSE function of $^4$He \cite{mazz3}.
When $q$ increases $R(q,Y)$ narrows and sharpens, 
becoming a delta function in the $q \rightarrow \infty$ limit.


The $^3$He additive correction $\Delta J(q,Y)= (q/m) \Delta S(q,Y)$ at
the same density and momentum transfer is also shown in Fig. \ref{fig-rqy}. This
function, which introduces  fermionic correlations to
the FSE, presents a shape  that is entirely different from that of $R(q,Y)$.
The strong oscillations that appear in the region $Y\approx\pm k_{\text F}$
modify the shape of the
IA response around these points. Furthermore, a central peak centered at
$Y=0$ enhances the strength of the total response at the origin. 
Out of this region ($|Y| \agt k_{\text F}$), $\Delta J(q,Y)$  is much
smaller and its correction to the response becomes negligible.
Further analysis
indicates that $\Delta J(q,Y)$ decays to zero in the high momentum
transfer limit. This fact, together with the limiting condition
$R(q\to\infty,Y)\to\delta(Y)$, indicates that the total response
asymptotically approaches $J(Y)$  when $q \rightarrow \infty$.

\begin{figure}
\centerline{
\includegraphics*[width=0.8\linewidth]{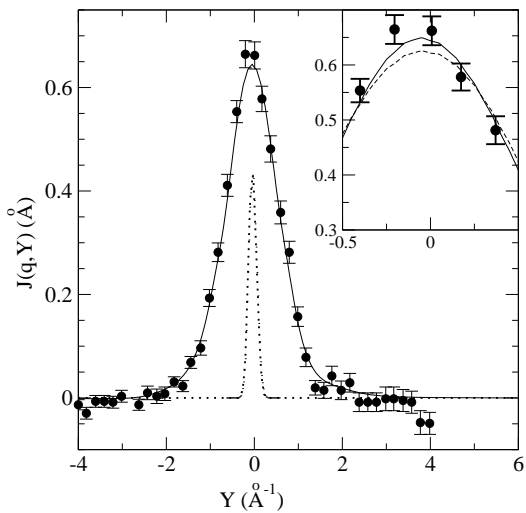}
}
\caption{$^3$He response at q=19.4 \AA$^{-1}$ and $\rho_0$ 
(solid line), folded to the IRE function (dotted line).  
The points with error bars are the experimental data from Ref. \protect\cite{azuah}. The instrumental
resolution function \protect\cite{azuah} has been divided by a 
factor 10 to fit into the scale. The inset shows $J(q,Y)$ near the peak 
with (solid line) and without (dashed line) the additive correction $\Delta
J(q,Y)$.}
\label{fig-tot}
\end{figure}

The final result of the $^3$He response at $q=19.4$ \AA$^{-1}$  is 
shown in Fig.~\ref{fig-tot}.
We compare our results with the DINS data of Azuah \textit{et al.}~\cite{azuah}
because their data correspond to densities around the equilibrium density
$\rho_0$, and also because their experimental setup produces a rather 
narrow instrumental resolution function. More recent data
\cite{senesi}  are focused
at higher liquid densities  and to the solid phase. In this experiment, the
momentum transfer is much larger ($q \sim 90$ to 120 \AA$^{-1}$) producing a
simultaneous decrease of the FSE corrections and a widening of the IRE. 
The IRE function estimated by Azuah \textit{et al.} \cite{azuah} is shown
in Fig. \ref{fig-tot} scaled by a factor 0.1. The theoretical response
$J(q,Y)$, shown in the figure, has been folded with the experimental IRE function
$I(q,Y)$ to make a direct comparison possible.
The kinks of $J(q,Y)$ at $Y=\pm k_{\text F}$ present in the IA, are completely 
washed out by the 
succesive folding with  $R(q,Y)$ and  $I(q,Y)$,
although $\Delta J(q,Y)$ introduces additional
structure around those points. The most remarkable feature 
is the enhancement of the strength of the response around the peak due to the
new addititve term introduced in the present approach (see inset in Fig.
\ref{fig-tot}). This small but significant increase of strength allows for
the first time to reproduce the available $^3$He DINS experimental data.

To summarize,    
we have presented a FSE formalism suitable to study the dynamic 
structure function of a Fermi liquid like liquid $^3$He at large momentum transfer. 
The method is a  natural extension of the Gersch-Rodriguez theory for
bosons. According to the present formulation, $J(q,Y)$ results from
the convolution of the IA with a purely 
bosonic FSE broadening function, which incorporates the  
 short-range correlations induced by the interatomic potential, 
plus an additive correction term that takes into
account Fermi statistics effects in the FSE. The results obtained are in
good agreement with DINS data, comparable for the first time to the accuracy
previously achieved in the study of the high-$q$ response of liquid $^4$He.
The two key features behind the present results are, on one hand, the
use of a realistic momentum distribution provided by the DMC method, and on
the other, the explicit introduction of Fermi corrections in the FSE. The
latter effect is estimated at the lowest order but its inclusion 
allows for a significant improvement and a better knowledge of 
specific mechanisms influencing the FSE in liquid $^3$He.

This work has been partially supported by
Grants No. BFM2002-01868  and BFM2002-00466 from DGI (Spain) and
Grants No. 2001SGR-00064  and 2001SGR-00222 from the Generalitat de Catalunya.

\end{document}